\documentclass[aps,preprint]{revtex4}
\usepackage{amsfonts}
\usepackage{amsmath}
\usepackage{amssymb}
\usepackage{graphicx}
\usepackage{pdfpages}
\usepackage{xcolor}
\usepackage{caption}
\usepackage{lineno,hyperref}
\usepackage{changes}%
\hypersetup{colorlinks =true, allcolors = blue}
\providecommand{\U}[1]{\protect\rule{.1in}{.1in}}

\begin{document}
\preprint{HEP/123-qed}
\title{Superradiant (In)stability, Greybody Radiation, and Quasinormal Modes of Rotating Black Holes in non-linear Maxwell $f(R)$ Gravity}
\author{Sara Kanzi}
\email{sara.kanzi@final.edu.tr}
\affiliation{Faculty of Engineering, Final International University, Kyrenia, North Cyprus via Mersin 10, Turkey}
\author{}
\affiliation{}
\author{\.{I}zzet Sakall{\i}}
\email{izzet.sakalli@emu.edu.tr}
\affiliation{Physics Department, Eastern Mediterranean
University, Famagusta, 99628 North Cyprus via Mersin 10, Turkey}
\author{Behnam Pourhassan}
\email{b.pourhassan@du.ac.ir}
\affiliation{School of Physics, Damghan University, Damghan, 3671641167, Iran.}

\begin{abstract}
The research of superradiant instability in the realm of quantum gravity is a well-known topic, with many physicists and astronomers studying the potential impact it can have on gravitational waves, the structure of the universe, and spacetime itself. In this work, we investigate the superradiant (in)stability of a rotating black hole obtained from the nonlinear Maxwell $f(R)$ gravity theory. In this study, the evaluation of stability/instability is going to be based on the non-existence and existence of the magnetic field, when the magnetic field constant becomes $c_{4}=0$ and $c_{4}\neq 0$, respectively. The analyzes of greybody factor (GF) and quasinormal modes (QNMs) are investigated in the stationary black hole spacetime both in the absence and presence of the magnetic field parameter. To this end, we first consider the Klein-Gordon equation for the complex scalar field in the geometry of that rotating black hole. In the sequel, the obtained radial equation is reduced to a one-dimensional Schr\"{o}dinger-like wave equation with effective potential energy. The effects of the nonlinear Maxwell $f(R)$ gravity theory parameters ($q$, $c$, and $c_{4}$) on the effective potential, GFs, and QNMs are thoroughly investigated. The obtained results show that even though the factors $q$, $c$, and $c_{4}$ all affect the effective potential, this phenomena, surprisingly, is not valid for the GFs and QNMs. With the proper graphics and tables, all outputs are depicted, tabulated, and interpreted.

\end{abstract}
\volumeyear{ }
\eid{ }
\date{\today}
\received{}

\maketitle
\tableofcontents

\section{Introduction}

Superradiance is a term used to describe a radiation amplification system that includes a scattering mechanism. Superradiance plays a noteworthy role in the studies of relativity, astrophysics, quantum mechanics, and optics \cite{s1}. This phenomenon can be considered as a quantum aspect of black hole (BH) physics. Namely, in the context of \textit{quantum gravity phenomenology}, superradiance can play a crucial role in the study of BHs. For example, the emission of Hawking radiation (or greybody radiation) \cite{hw1,hw2} from a BH can be enhanced by the presence of surrounding matter, leading to a process known as BH superradiance. This phenomenon has been proposed as a possible explanation for the \textit{observed properties of BH systems}, such as the large amounts of energy emitted from the centers of galaxies. Because it is influenced by the BH's area theorem, tidal forces, the Penrose process, and Hawking radiation \cite{McCaughey:2016nhr}. The concept of superradiance was first introduced by Dicke in 1954 \cite{1}. Subsequently, in 1971, this phenomenon was further understood and developed by Zel'dovich \cite{2,3} during his investigation of reflected wave amplification from a rotating BH (Kerr metric). Zel'dovich proved that if the frequency $\omega$ of the ingoing wave having the plane wave structure with $e^{-i\Omega t+im\phi}$ ($\Omega$ represents the angular velocity, $\phi$ is the cyclic coordinate, and $m$ denotes the magnetic quantum number) convinces $\omega< m\Omega$, the scattered waves amplify in such a way that the waves coming out of the BH to become more than the ones entering to it. Namely, the dispersed wave is amplified.

Combination of superradiance with a confining procedure to force the wave to
consistently interact with the BH to undergo exponential increase known as "BH bomb" \cite{4}, which has been an active case since 1970s \cite{5}. This phenomenon can be viable by considering a mirror around the rotating body that could make the system unstable \cite{3,Cardoso:2004nk}. In \cite{s5}, it was mentioned that such a mirror
can be recognized by applying a charged massive scalar field breeding in the
Kerr-Newman spacetime. Moreover, BHs can be transformed into efficient particle detectors by applying strong constraints to ultralight bosons via the superradiant instabilities of spinning BHs. On the other hand, instability formation and whether or not its nonlinear time evolution follows the linear intuition are, nevertheless, topics that are not well-understood, yet \cite{Brito:2014wla}. 

A Kerr BH might be thought as a strong candidate for superradiant phenomena
\cite{6,7,8}, however BHs in Kerr form do not exhibit superradiant instabilities with significant growth rates \cite{Cardoso:2004nk,10,11,12}. In fact, there are various methods for using superradiance to extract energy from BHs: 1) BH fission \cite{Brito:2015oca}, 2) BH bombs (as mentioned above) \cite{4,Cardoso:2004nk}, 3) Accretion disks and torus \cite{Bekenstein:1998nt}, 4) BH bombs in anti-de Sitter (AdS) spacetime \cite{Cardoso:2004hs,Hawking:1999dp}, 5) Massive fields, soft bombs, and particle physics \cite{Cardoso:2013krh,Pani:2012vp,Arvanitaki:2010sy}, 6) Floating orbits \cite{Kapadia:2013kf}, 7) Generalized scalar-tensor theories and superradiance \cite{Cardoso:2013krh,Cardoso:2013opa}, 8) Ergoregion instability \cite{Friedman,Moschidis:2016zjy}. On the other hand, today, there are various theories for modified gravity, such as brane-world gravity
\cite{22}, Dvali-Gabadadze-Porrati gravity \cite{23}, Einstein-Aether theory
\cite{24,JHEP2}, tensor-vector-scalar theory \cite{25}, and $f\left(  R\right)  $
gravity \cite{26,27,28,29}, which all attracted much attention in the literature. $f(R)$ theories of
gravity are straightforwardly generated by replacing Ricci scalar $R$ in the
Einstein-Hilbert action. Namely, we have a generic action for $f(R)$ as follows %
\begin{equation}
S=\frac{1}{2k}\int d^{4}x\sqrt{-g}f\left(  R\right)  , \label{1}%
\end{equation}
where $k=8\pi$ denotes the gravitational constant and $g$ is the
determinant of metric. Throughout the paper, unless otherwise noted, we shall work in natural units with $c_{s}=G=\hbar=1$. Since general relativity (GR) has had many unresolved problems, including the existence of dark energy and dark matter, deflection from Einstein's theory allows us to estimate the fundamental matters and extension of GR (modified gravity). Based on different formalisms, there are three types of $f\left(
R\right)  $ gravity models: 1) metric, 2) Palatini, and 3) metric-affine, $f\left(  R\right)$
gravity \cite{30,31,32}. The use of $f (R)$ gravity in many contexts is significant; for example, see its astrophysical perspective in \cite{33,34,35,36}, the cosmological models with $f\left(  R\right)$ in \cite{37,38}, and the derivations of novel BH solutions in \cite{39,40,41,42}. Moreover, we
refer the reader to the monographs \cite{43,44} for some good reviews about $f\left(  R\right)$. In particular, spherically symmetric BH
solutions in $f\left(  R\right)$ gravity have been receiving special attentions (see for instance \cite{45},
in which the exact static spherically symmetric solutions in $f\left(
R\right)$ gravity coupled with nonlinear electrodynamics derived by
Hollestein and Lobo \cite{46}). Searching for alternative gravitational theories to conventional Einstein's general relativity (GR) is supported by difficult challenges ranging from quantum gravity to dark energy and dark matter. Indeed, there are a lot of unresolved problems with GR, such as singularities, the nature of dark energy and dark matter. All of these problems motivate researchers to improve or modify GR in order to address the challenges at the UV and IR scales \cite{Capozziello:2011et}. However, the obtained feasible modified/extended theories should be consistent with the present observational/experimental restrictions. With the awareness of this issue, an ambitious study has recently been carried out on the nonlinear Maxwell $f(\mathcal{R})$ gravity \cite{45}. By using dynamical Ricci scalars that asymptotically converge to flat or (A)dS spacetimes, Nashed and Saridakis \cite{72} have derived a new charged rotating BH solutions, which will be our main reference metrics in this study.

Due to the quantum effects, a BH can act as a blackbody object, which emits
thermal waves \cite{hw1,hw2}. The mass of a BH decreases during
its HR, which can lead to complete BH evaporation. As a matter of course, the emitted
particles are affected by an effective potential originating from the curvature of the spacetime. As a result, although some waves penetrate the potential barrier and extend to infinity, the remainder is reflected back to the BH. Due to the structure of the effective potential, the radiation spectra are altered and different from that near the event horizon. As a result, the term GF \cite{Sakalli:2022xrb} refers to a quantity that measures the deviation of the radiation spectrum from the blackbody radiation. At the event horizon, the BH emission rate \cite{Rocha:2009xy} is defined as follows 
\begin{equation}
\gamma\left(  \omega\right)  =\left(  \frac{d^{3}k}{8\pi^{3}e^{\frac{\omega
}{T_{H}}}}\right)  , \label{2}%
\end{equation}

by which $\omega$ represents the wave frequency, $T_{H}$ and $k$ denote the Hawking temperature
and surface gravity, respectively. The relation between emission rate and GF is given by \cite{50}
\begin{equation}
\gamma\left(  \omega\right)  =\left(  \frac{d^{3}k\left\vert A_{l,m}%
\right\vert ^{2}}{8\pi^{3}e^{\frac{\omega}{T_{H}}}}\right)  , \label{3}%
\end{equation}

where $\left\vert A_{l,m}\right\vert ^{2}$ represents the GF. There are
various methods to compute the GF, such as matching technique \cite{49, 50,
51}, WKB approximation \cite{52, 53}, finding Bogoliubov coefficients
method \cite{54,55,56,57}, the Miller--Good transformation method
\cite{58,Sakalli:2022xrb}, and the rigorous bounds \cite{59}.

Teukolsky equation \cite{60} describes an oscillation system that naturally dissipates. Such a system generates QNMs instead of the classical normal mode solution. Vishveshwara was the first to identify the QNMs of a BH
\cite{s60}. The QNMs are described by complex frequencies that carry characteristic information about the BH spacetime, which is in the ringdown phase. The QNMs have broad literature in the
BH physics. Specifically, explicit superpositions of QNMs may be utilized to estimate the gravitational wave frequencies in the gravitational wave phenomenon \cite{61,62,63,64}. There are many rich and excellent investigations on the QNMs of various solutions of BHs \cite{68,69,70,71,JHAP}, which are considered seminal works of the subject
\cite{65,66,67}.

The paper is divided into the following sections. In Sec. \eqref{sec2}, we introduce the metric of the rotating BH in nonlinear Maxwell $f\left( R\right)$ gravity and demonstrate some of its physical features. Section \eqref{sec3} is
devoted to the Klein-Gordon equation (KGE) for charged massive scalar fields in that
rotating BH geometry. In this section, we show that the radial wave equation reduces a one-dimensional Schr\"{o}dinger-like wave equation with a corresponding effective potential. We also study the behaviors of the obtained effective potential under the influence of charge $q$ and magnetic field constant $c_{4}$. As being two subsections, in Sec. \eqref{sec4}, we examine the superradiant instability for zero and non-zero $c_{4}$ values. Sections \eqref{sec5} and \eqref{sec6} are reserved for the analysis of the greybody radiation and QNMs,
respectively. Our results are summarized and discussed in Sec. \eqref{sec7}. We follow the metric convention $(+,-,-,-)$.

\section{Rotating BH\lowercase{s} in Nonlinear Maxwell $f\left(  R\right)$ gravity} \label{sec2}%

\begin{figure}[h]
\centering\includegraphics[width=9cm,height=10cm]{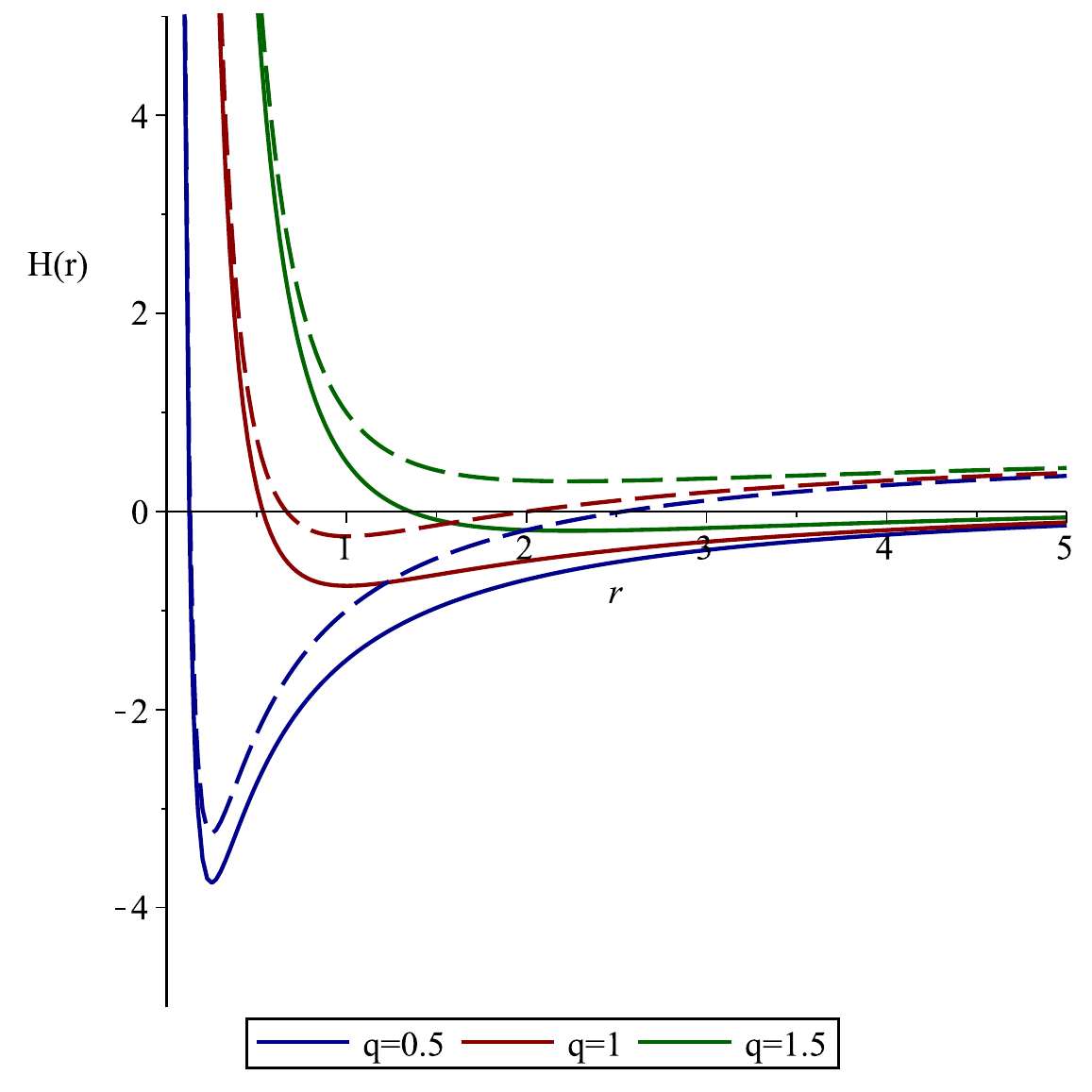}\caption{Schematic plots
of $H(r)$ versus $r$. Solid lines represent $c=0.5$ and dash lines are for $c=1.5.$ The physical parameters are chosen as $M=1$ and $a=0.3$.} \label{H1}
\end{figure}

In this section, we briefly review both new static and rotating BH solutions
obtained in nonlinear Maxwell $f\left(  R\right)  $ gravity \cite{72}, whose the total action is given by%

\begin{equation}
S_{t}=\frac{1}{2k}\int\sqrt{-g}f(R)d^{4}x+\int\sqrt{-g}%
\mathcal{L}%
(%
\mathcal{F}%
)d^{4}x, \label{4}%
\end{equation}
where $k$ is a gravitational constant, which can be considered as $k=1$, without loss of generality, in this study. \textcolor{blue}{Moreover, $\mathcal{L}(\mathcal{F})$ indicates  a general gauge-invariant electromagnetic Lagrangian where the usual antisymmetric Faraday tensor is $\mathcal{F}=\frac{1}{4}\mathcal{F}_{\alpha\beta}\mathcal{F}^{\alpha\beta}$.}
$\sqrt{-g}$ represents the determinant of the metric $g_{\mu\nu}.$ The corresponding gravitational field equations of the action \eqref{4} can be derived as follows%
\begin{equation}
\textcolor{blue}{\xi_{\mu\nu}=R_{\mu\nu}f_{R}-\frac{1}{2}g_{\mu\nu}f(R)-2g_{\mu\nu
}\Lambda+g_{\mu\nu}\nabla_{\alpha}\nabla^{\alpha}f_{R}-\nabla_{\mu}%
\nabla_{\nu}f_{R}-8\pi\Im_{\mu\upsilon}^{nlem}\equiv0}, \label{5}%
\end{equation}
by which $\Im_{\mu\upsilon}^{nlem}$ denotes the energy momentum tensor and
$f_{\Re}\equiv\frac{df(\Re)}{d\Re}$. Using the following ansatz for a spherically symmetric line element: 

\begin{equation}
ds^{2}=H\left(  r\right)  dt^{2}-\frac{dr^{2}}{H\left(  r\right)  }%
-r^{2}\left(  d\theta^{2}+\sin^{2}\theta d\varphi^{2}\right)  , \label{6}%
\end{equation}
in Eq. \eqref{5}, after making some tedious calculations, the following metric function was finally obtained by Nashed and Saridakis \cite{72}
\begin{equation}
H(r)=\frac{c}{2}-\frac{2M}{r}+\frac{q^{2}}{r^{2}}, \label{7}%
\end{equation}
where $c$ is a positive constant, which can take limited values: $0< c\leq 2$, $q$ and $M$ stand for the charge and mass, respectively (see Ref. \cite{72} for the details). In Fig. \ref{H1}, we show the behavior of the metric function $H\left(  r\right)$ under the influence of varying parameters $q$ and $c$. It is clearly seen that the spacetimes exhibit flatness at the asymptotic distances, independent of the values of $q$ and $c$. In that case, the horizon radius is obtained as,
\begin{equation}
r_{h}=\frac{2M}{c}\left(1+\sqrt{1-\frac{cq^{2}}{2M^{2}}}\right). \label{horizon}%
\end{equation}
Hence, it will be Schwarzschild black hole if we set $c=2$ and $q=0$. Also, extremal limit is given by the following condition,
\begin{equation}
cq^{2}=2M^{2}. \label{extremal}%
\end{equation}

The rotating version of the BH solution can
be derived by applying the following transformations \cite{Lemos:1994xp,Awad:2019jur}
\begin{align}
\overset{\thicksim}{\phi}  &  =\Xi\phi+at,\nonumber\\
\overset{\thicksim}{t}  &  =\Xi t+a\phi, \label{s9}%
\end{align}
to the static and spherically symmetric metric \eqref{6}. In Eq. \eqref{s9}, $a$ is the rotating parameter and $\Xi=\sqrt{1+a^{2}}$.  Thus, we have%
\begin{multline}
ds^{2}=[\Xi^{2}H(r)-a^{2}r^{2}\sin^{2}\theta]dt^{2}-\frac{dr^{2}}{H(r)}%
-r^{2}d\theta^{2}-\\
\lbrack\Xi^{2}r^{2}\sin^{2}\theta-a^{2}H(r)]d\phi^{2}+2a\Xi\lbrack
H(r)-r^{2}\sin^{2}\theta]dtd\phi, \label{10}%
\end{multline}
in which $H(r)$ is provided by the static solution \eqref{7} previously derived. On the other hand, the general gauge potential is defined by \cite{72}
\begin{equation}
\overset{\thicksim}{V}=[\Xi q(r)+as(r)]d\overset{\thicksim}{t}+n(\phi
)dr+[aq(r)+\Xi s(r)]d\overset{\thicksim}{\phi}, \label{11}%
\end{equation}
where $q(r)$, $s(r)$, and $n(\phi)$ are 3 free functions generating the electric and magnetic charges in the vector potential as follows
\begin{equation}
    s(r)=c_{4}r,\text{ \ \ \ \ \ \ \ } n(\phi)=c_{4}\phi, \label{s11}%
\end{equation}
in which $c_{4}$ represents the magnetic field constant. In the following sections, our investigation will consider cases in which the magnetic field constant does not exist ($c_{4}=0$) and exists ($c_{4}\neq 0$).  

\section{Scalar Perturbation}   \label{sec3}

In recent decades, perturbations of BHs and stars have arisen as one of the main topics of relativistic astrophysics. Furthermore, perturbations are hot subjects right now because of their functions in gravitational waves. In this part, we employ the charged KGE to arrive at the Schr\"{o}dinger wave equation in one dimension. The effective potential to be obtained in this section is crucial for studying superradiance, greybody radiation, and QNMs.

Let us consider the charged and massive KGE:

\begin{equation}
\frac{1}{\sqrt{-g}}\left(  \partial_{\mu}-iQA_{\mu}\right)  \left(  \sqrt
{-g}g^{\mu\nu}\left(  \partial_{\mu}-iQA_{\nu}\right)  \Psi\right)  =m^{2}\Psi,   \label{12}%
\end{equation}
where $Q$ and $m$ are the charge and mass of the scalar field (spin-$0$), respectively.
Moreover, $\sqrt{-g}$ represents the determinant of the metric. Here, for metric (\ref{10}), we consider the following ansatz for the spinor field:
\begin{equation}
\Psi=e^{-i\omega t}e^{ik\phi}R\left(  r\right)  Y\left(  \theta\right),
\label{13}%
\end{equation}

\textcolor{blue}{where $\omega$ represents the frequency of the wave and $k$ is azimuthal quantum number}. During the derivation of the scalar wave equation, we will consider the dyonic case. Plus, we set $s(r)=n(\phi)=0$ in Eq. (\ref{11}).  Thus, the components of the vector potential read%
\begin{equation}
A_{\overset{\thicksim}{t}}=\Xi q,\text{ \ \ \ and \ \ \ }  \color{black}{A_{\overset{\thicksim
}{\phi}}=aq}. \label{14}%
\end{equation}
Throughout the paper \textcolor{blue}{for more convenience in our calculation}, without loss of generality, we shall consider $qQ\rightarrow q^{2}$. After substituting Eq. \eqref{14} and ansatz (\ref{13}) into the massive charged KGE (\ref{12}), one can obtain

\begin{multline}
\left(  \frac{2H}{r}+H^{\prime}\right)  R^{\prime}\left(  r\right)
+HR^{\prime\prime}\left(  r\right)  +\\ 
\frac{1}{H}\left[  \Xi^{2}\omega^{2}+\Xi^{4}q^{4}+2q^{2}\omega\Xi^{3}%
+2a\omega\Xi k+2a\Xi^{2}q^{2}k-2a^{2}\Xi^{2}q^{4}-\right. \\ 
\left.  2\omega a^{2}q^{2}\Xi+a^{2}k^{2}+a^{4}q^{4}-2kq^{2}a^{3}+m^{2}%
+\lambda\right]  R\left(  r\right)  =-\lambda, \label{15}%
\end{multline}

where $\lambda$ is the eigenvalue whose value can be found with the help of the angular part:
\begin{equation}
\cot\theta Y_{\theta}\left(  \theta\right)  +Y_{\theta\theta}\left(
\theta\right)  -\frac{\left(  a\omega+\Xi k\right)  ^{2}}{\sin^{2}\theta}Y(\theta)=0.
\label{16}%
\end{equation}

In the mean time, throughout the paper, a prime (dash) symbol is used to denote the derivative of a function with respect to its argument. By considering the definition of the tortoise coordinate:
\begin{equation}
r_{\ast}=\int\frac{dr}{H\left(  r\right)  }, \label{s16}%
\end{equation}
and in the sequel applying the transformation $R(r)=\frac{U\left(  r\right)  }{r}$ to Eq. \eqref{16}, one can acquire one-dimensional
Schr\"{o}dinger like wave equation as follows:
\begin{equation}
\frac{d^{2}U\left(  r\right)  }{dr_{\ast}^{2}}+\left\{  \varpi^{2}%
-V_{eff}\right\}  U\left(  r\right)  =0, \label{17}%
\end{equation}
in which $\varpi^{2}=\omega^{2}\left(  1+a^{2}\right)
=\omega^{2}\Xi^{2}$ and the effective potential is given by%
\begin{equation}
V_{eff}=-2\Xi\left(  q^{2}+ak\right)  \omega-\left(  q^{2}+ak\right)
^{2}+\frac{HH^{\prime}}{r}+\frac{\lambda H}{r^{2}}+m^{2}H. \label{18}%
\end{equation}

\begin{figure}[h]
\centering\includegraphics[width=9cm,height=10cm]{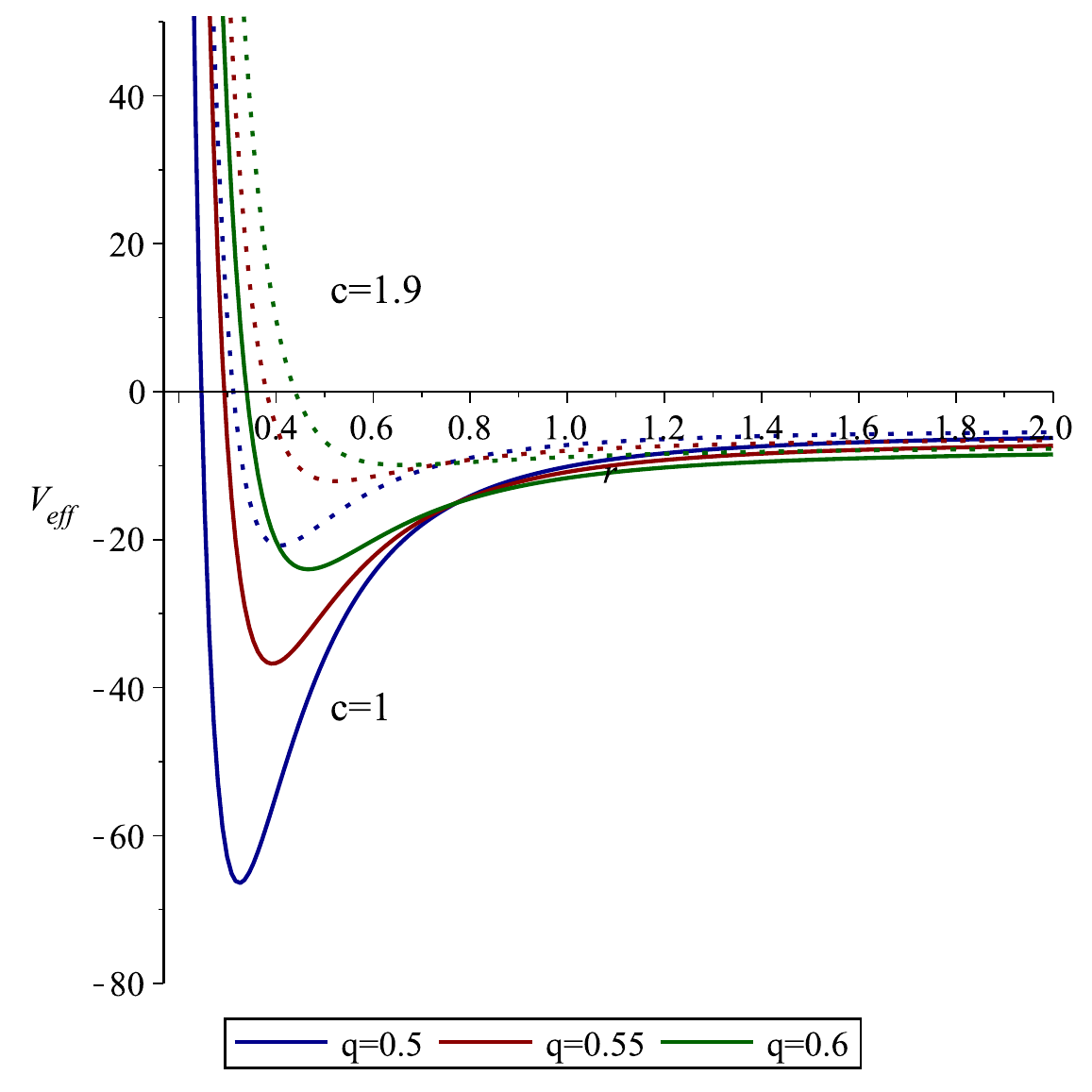}\caption{Plots of
$V_{eff}$ versus $r$ for the spin-0 particles in the case of zero magnetic  constant. The physical parameters are
chosen as; $M=m=1,\omega=15,a=0.3,$ and $\lambda=2$.}  \label{F2}
\end{figure}

The behaviors of the effective potential \eqref{18} when the charge parameter $q$ is changed for various values of $c$ are depicted in Fig. \ref{F2}.

\section{Superradiance Phenomenon} \label{sec4}

\subsection{For \texorpdfstring{$c_{4}=0$}{c4=0}}
Here, we investigate the stability of the rotating BH obtained from the non-linear Maxwell $f\left(  R\right)  $ gravity. To this end, we consider the method prescribed in Ref.
\cite{s73,s74}. After applying the transformation $U\left(  r\right)  =e^{-i\varpi
r_{\ast}}\psi\left(  r\right)  $ to the Schr\"{o}dinger equation seen in Eq. (\ref{17}), one gets%
\begin{equation}
\frac{d^{2}\psi\left(  r\right)  }{dr_{\ast}^{2}}-2i\varpi\frac{d\psi\left(
r\right)  }{dr_{\ast}}-V_{eff}\psi\left(  r\right)  =0. \label{19}%
\end{equation}
Now, let us replace the tortoise coordinate (\ref{s16}) with the naive radial coordinate $r$
\begin{equation}
\frac{d}{dr}\left(  H\frac{d\psi}{dr}\right)  -2i\varpi\frac{d\psi}{dr}%
-\frac{V_{eff}}{H}\psi=0, \label{20}%
\end{equation}
and multiply Eq. (\ref{20}) by $\psi^{\ast}$. By imposing $H\left(  r_{+}\right)  =0$ and $\psi\left(  \infty\right)  =0$, one can solve the final differential equation by performing the well-known integration by parts method. Thus, we get
\begin{equation}
\int_{r_{h}}^{\infty}dr\left[  H\left\vert \frac{d\psi}{dr}\right\vert
^{2}+2i\varpi\psi^{\ast}\frac{d\psi}{dr}+\frac{V_{eff}}{H}\left\vert
\psi\right\vert ^{2}\right]  =0, \label{21}%
\end{equation}
where the second term in the integrated can be expanded to
\begin{equation}
2i\varpi\frac{d\psi}{dr}=\varpi\psi^{\ast}\frac{d\psi}{dr}+\overset{-}{\varpi
}\psi\frac{d\psi^{\ast}}{dr}. \label{22}%
\end{equation}

Therefore, the integration (\ref{21}) recasts in
\begin{equation}
\int_{r_{h}}^{\infty}dr\left[  H\left\vert \frac{d\psi}{dr}\right\vert
^{2}+\frac{V_{eff}}{H}\left\vert \psi\right\vert ^{2}\right]  =-\frac
{\left\vert \varpi\right\vert ^{2}\left\vert \psi\left(  r_{h}\right)
\right\vert ^{2}}{\operatorname{Im}\omega}. \label{23}%
\end{equation}

It is also possible to write Eq. (\ref{23}) as 
\begin{equation}
\int_{r_{h}}^{\infty}dr\left[  H\left\vert \frac{d\psi}{dr}\right\vert
^{2}+\frac{\overset{\thicksim}{V}_{eff}}{H}\left\vert \psi\right\vert
^{2}-\frac{\left(  q^{2}+ak\right)  ^{2}}{H}\left\vert \psi\right\vert
^{2}-\frac{2\Xi\left(  q^{2}+ak\right)  \omega}{H}\left\vert \psi\right\vert
^{2}\right]  =-\frac{\left\vert \varpi\right\vert ^{2}\left\vert \psi\left(
r_{h}\right)  \right\vert ^{2}}{\operatorname{Im}\omega}. \label{24}%
\end{equation}

In Eq. \eqref{24}, $\overset{\thicksim}{V}_{eff}$ stands for the potential terms without $q$ parameter in  $V_{eff}$. In addition, it is discovered that the final term of Eq. (\ref{24}) has almost no impact on superradiance calculations and the sign of expression $\frac{\overset{\thicksim}%
{V}_{eff}}{H}\left\vert \psi\right\vert ^{2}-\frac{\left(  q^{2}+ak\right)
^{2}}{H}\left\vert \psi\right\vert ^{2}$ is critical for evaluating the
stability of the black hole. Meanwhile, the potential \eqref{18} is positive outside the horizon, which means $Im(\omega)$ must be negative. Thus, in light of the Dirichlet boundary conditions, one can conclude that the scalar field propagation will be stable.
In order to assess the superradiant instability
of the rotating black hole in non-linear Maxwell $f\left(  r\right)  $
gravity in a more authentic form, we shall define the reflection/transmission coefficients to determine the superradiant condition. To this end, we shall perform our computations in three different regions. The first region (Region I) is the region that is close to the event horizon $\left(  r\thickapprox
r_{h}\right)  ,$ where the potential is approximated to%
\begin{equation}
V_{eff}\approx-2\Xi\left(  q^{2}+ak\right)  \omega-\left(  q^{2}+ak\right)  ^{2},
\label{25}%
\end{equation}
and correspondingly
\begin{equation}
\overset{\thicksim}{V}_{eff}\ll\left(  \omega\Xi+\left(  q^{2}+ak\right)
\right)  ^{2}. \label{26}%
\end{equation}

Thus, in Region I, the solution of the radial equation \eqref{17} is obtained as
\begin{equation}
U_{1}\left(  r\right)  =Ae^{-i\left(  \omega\Xi+\left(  q^{2}+ak\right)  \right)
r_{\ast}}. \label{27}%
\end{equation}
which slightly away from the event horizon yields \cite{s74}
\begin{equation}
U_{1}\left(  r\right)  \approx A\left(  1-\frac{i\left(  \omega\Xi+\left(
q^{2}+ak\right)  \right)  }{H^{\prime}\left(  r_{h}\right)  }\ln\left(
r-r_{h}\right)  \right), \label{28}%
\end{equation}
where the near horizon tortoise coordinate is defined as $r_{\ast}=\int\frac{dr}{H\left(
r\right)  }\thickapprox\frac{1}{H^{\prime}\left(  r_{h}\right)}\ln\left(
r-r_{h}\right)$.

Between the event horizon and the distant regions, Region II acts as an intermediate zone and is defined as:
\begin{equation}
\overset{\thicksim}{V}_{eff}\gg\left(  \omega\Xi+\left(  q^{2}+ak\right)
\right)  ^{2}. \label{29}%
\end{equation}

Therefore, the radial equation \eqref{17} reads
\begin{equation}
\frac{d^{2}U}{dr_{\ast}^{2}}=0,\Rightarrow U\left(  r_{\ast}\right)  =B+C\int
dr_{\ast}. \label{30}%
\end{equation}
By comparing the solutions in the first and second regions, we define the constants as
as $A=B,$ and $C=-Ai\left(  \omega\Xi+\left(  q^{2}+ak\right)  \right)  $.

To find the solution in Region II, we take into consideration an asymptotic series for tortoise coordinate i.e., $r\gg r_{h}$. Hence, we get
\begin{equation}
U_{2}\left(  r\right)  =A\left(  1-\frac{i\left(  \omega\Xi+\left(
q^{2}+ak\right)  \right)  }{4r^{4}}\overset{\thicksim}{k}\right)  , \label{31}%
\end{equation}
where%
\begin{equation}
\overset{\thicksim}{k}=-\frac{32}{c^{2}r^{5}}\left\{  32M^{3}cq^{2}%
-3Mc^{2}q^{4}-64M^{5}\right\}  . \label{32}%
\end{equation}
The third region (Region III) is the asymptotic zone ($r\gg r_{h},$), where the conducting terms of the effective potential become
\begin{equation}
V_{eff}\approx -2\Xi\left(  q^{2}+ak\right)  \omega-\left(  q^{2}+ak\right)
^{2}+\frac{m^{2}c}{2}. \label{33}%
\end{equation}
Thus, one can get the Region III solution as 
\begin{equation}
U_{3}\left(  r\right)  =D_{1}+D_{2}e^{-i\sqrt{\left(  \omega\Xi+\left(  q^{2}+ak\right)
\right)  ^{2}+\frac{m^{2}c}{2}}r_{\ast}}. \label{34}%
\end{equation}
Then, after matching the solution of Region II with the solution of Region III, we get
\begin{equation}
D_{2}=Ae^{\frac{i\overset{\thicksim}{k}}{4r^{4}}\eta},
\label{35}%
\end{equation}
where
\begin{equation}
    \eta=\frac{m^2c}{4(\omega\Sigma+(q^2+ak))}. \label{S35}
\end{equation}
To obtain the reflection coefficient and the GFs and to complete our assessment of superradiant stability/instability, we employ the flux expression as follows
\begin{equation}
    F=\frac{\sqrt{-g}g^{rr}}{2i}\left(U^{\ast}\partial_{r} U-U\partial_{r}U^{\ast}\right). \label{ss34}
\end{equation}
Therefore, one can obtain the near-horizon flux as 
\begin{equation}
    F_{hor}\propto -A^{2}\left(\omega\Xi+\left(q^{2}+ak\right)\right).
\end{equation}
Similarly, the asymptotic flux at spatial infinity becomes
\begin{equation}
    F_{\infty}\propto -\xi\left[1+\frac{D_{1}}{2}\left(D_{2}e^{-i\xi }+D_{2}^{\ast}e^{i\xi }\right)\right], \label{ss35}
\end{equation}
where $\xi=\sqrt{\left(\omega\Xi+(q^2+ak)\right)^2+\frac{m^2c}{2}}.$ By considering $D_{1}=\hat{D_{1}}+\hat{D_{2}}$ and $D_{2}=i(\hat{D_{2}}-\hat{D_{1}})$, the asymptotic incoming and
outgoing fluxes \eqref{ss35} can be written as follows
\begin{equation}
    F_{\infty-in}\propto -\xi\left[1-i\mid D_{2}\mid ^{2}sinh(i\xi )\right], \label{ss36}
\end{equation}
and
\begin{equation}
    F_{\infty-out}\propto -\xi\left[1+i\mid D_{1}\mid^{2}sinh(i\xi )\right]. \label{ss37}
\end{equation}
By employing the definition of the reflection coefficient and GF, we get 
\begin{equation}
    R=\mid\frac{F_{\infty-out}}{F_{\infty-in}}\mid=\left(\frac{1+iA_{1}^{2}sinh(i\xi )}{1-iA_{1}^{2}sinh(i\xi )}\right), \label{ss38}
\end{equation}
and 

\begin{equation}
    \gamma=\frac{F_{hor}}{F_{\infty-in}}=\frac{A^2\left(\omega\xi+(q^2+ak)\right)}{\xi\left(1-iA_{1}^{2}sinh(i\xi )\right)}. \label{ss39}
\end{equation}
Now, based on Eqs. \eqref{ss38} and \eqref{ss39}, we are able to determine the superradiant condition. To this end, either the reflection coefficient should be greater than 1 or the GF should be negative:
\begin{equation}
    \omega\leq\frac{-(q^2+ak)}{\xi}. \label{ss40}
\end{equation}
By taking into account superradiant instability conditions, in the relevant sub-cases, we now review the behavior of the effective potential in Fig. \ref{F2}. It is obvious from Fig. \ref{F2} that the potential does not contain wells and hence there are no bound states, which can prevent the accumulation of energy that might cause the instability. This indicates that the rotating BH in non-linear Maxwell $f(R)$ gravity can readily absorb the charged scalar wave and whence the associated background becomes stable under the charged scalar perturbations.

\subsection{For Infinitesimal $c_{4}$}
In this sub-section, our aim is to evaluate the superradiant stability/instability of the stationary BH found in the non-linear Maxwell $f(R)$ gravity with the case of $c_{4}$. However, we shall choose the magnetic field constant to be infinitesimally small to facilitate calculations. Moreover, we shall determine the effective potential with the aid of Eq. \eqref{19}, not from the Schr\"{o}dinger equation as it was done in the previous sub-section. So, following the steps we did earlier, we get 
\begin{equation}
    \frac{d}{dr}(H(r)\frac{d\Psi}{dr})-2i\Bar{\omega}\frac{d\Psi}{dr}-\frac{V_{eff}}{H(r)}\Psi=0, \label{K40}
\end{equation}
where $\Bar{\omega}=H(\omega+qc_{4}\phi)$ and the effective potential is determined as a complex expression. In Region 3, its real part can be expressed as
\begin{equation}
    Re[V_{eff}]\approx -H^2\omega^2+\frac{HH^{\prime}}{r}-2qc_{4}\phi H^2\omega+(\omega\Xi+(q^2+ak))^2-H(\frac{\lambda}{r^2}-m^2),  \label{K41}
\end{equation}
and its imaginary part reads
\begin{equation}
    Im[V_{eff}]\approx -\omega H^{\prime}H.  \label{K42}
\end{equation}

\begin{figure}[h]
\centering\includegraphics[width=9cm,height=10cm]{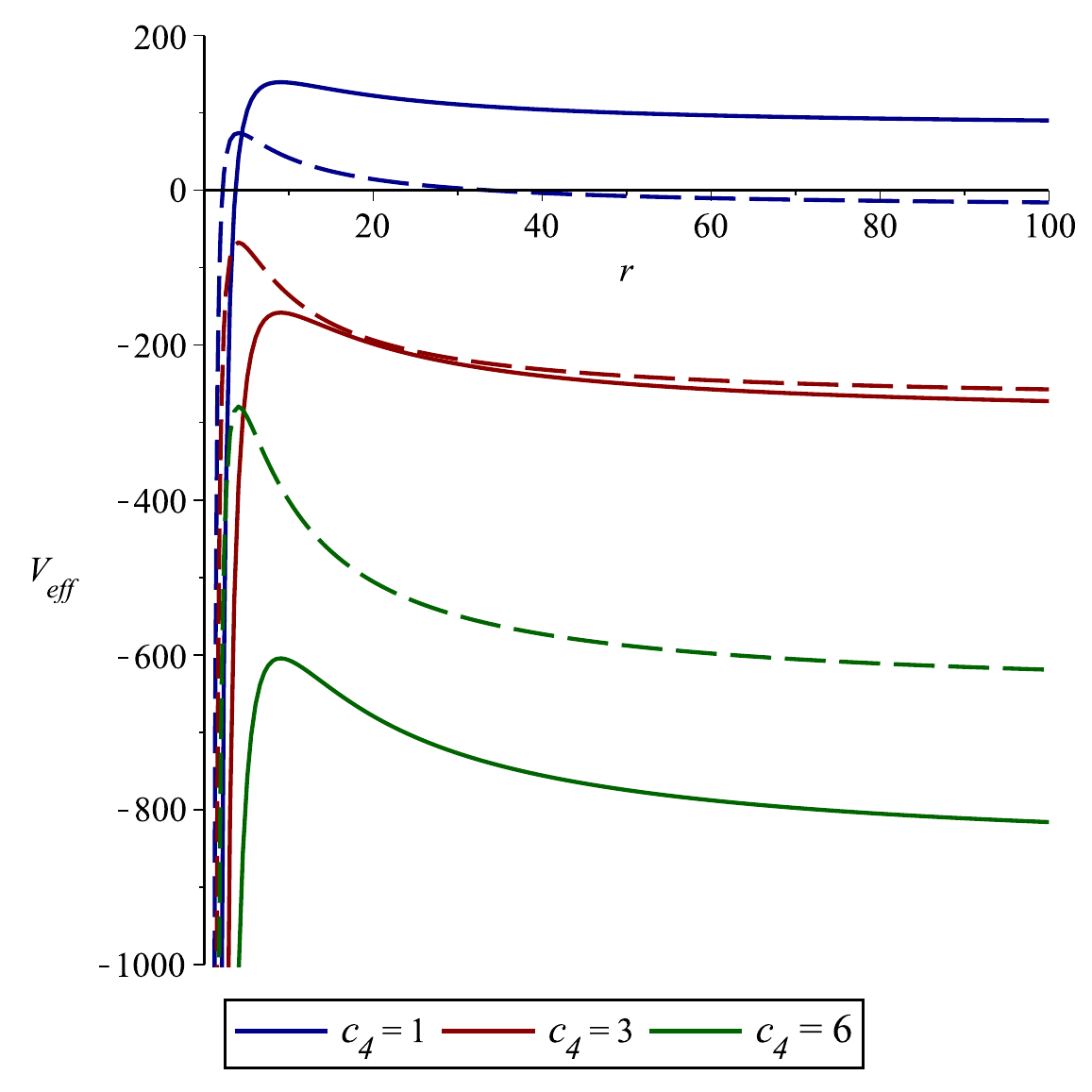}\caption{Plots of $V_{eff}$ versus $r$ for the spin-0 particles and non-zero $c_{4,}$. The solid lines represent the effective potential for $q=3$ and the dashed ones stand for $q=2$. The physical parameters are chosen as; $M=m=c=1,\omega=10,a=0.1,$ and $\lambda=2$.}  \label{Fs2}
\end{figure}

In Eqs. \eqref{K41} and \eqref{K42}, the terms including $c_{4}^{2}$ and $\frac{c_{4}}{r^2}$ are ignored. In addition, our analysis has shown us that it is reasonable to consider only the real component of the effective potential.\\

The wave solutions of regions 1 and 2 for the existing magnetic fields of the BH are the same as the non-existing ones, Eqs. \eqref{27} and \eqref{31}, but the wave solution of Region 3 $(r>>r_{h})$ with $c_{4}\neq 0$ is different than the $c_{4}=0$ one:
\begin{equation}
    U_{3}(r)=D_{1}+D_{2}\exp\left(-i\sqrt{V_{eff(3)}}\frac{\overset{\thicksim}{k}}{{4r^4}}\right)r_{\ast}, \label{K43}
\end{equation}
where
\begin{equation}
 V_{eff(3)}=-\frac{\omega^2 c^4}{4}-\frac{1}{2}(qc_{4}\phi c^2\omega-m^2c)+(\omega\Xi+(q^2+ak))^2, \label{K44}   
\end{equation}
and $\overset{\thicksim}{k}$ is nothing but Eq. \eqref{26}. Comparing Eq. \eqref{K43} with the solution obtained for Region 2 \big(Eq. \eqref{31}\big), one can determine the unknown constants as $D_{1}=A$ and 
\begin{equation}
    D_{2}=A\exp\bigg({\frac{i\overset{\thicksim}{k}}{4r^4}}\overset{\thicksim}{\eta}\bigg), \label{K45}
\end{equation}
in which 
\begin{equation}
   \overset{\thicksim}{\eta}=\frac{2m^2c-2qc_{4}\phi c^2\omega-\omega^2 c^4}{8\big(\omega\Xi+(q^2+ak)\big)}. \label{K46} 
\end{equation}
To determine the flux expressions at the horizon and spatial infinity, we apply the same method followed in the previous sub-section. Thus, we have
\begin{equation}
    F_{hor}\propto -(\omega\Xi+(q^2+ak))|A|^2, \label{K47}
\end{equation}
and
\begin{equation}
    F_{\infty}\propto -\left(1+\frac{D_{1}}{2}(D_{2}e^{-i\beta}+D_{2}^\ast e^{i\beta})\right), \label{K48}
\end{equation}
where 
\begin{equation}
   \beta=\sqrt{V_{eff(3)}}\frac{\overset{\thicksim}{k}}{4r^4}. \label{K49}
\end{equation}
By substituting $D_{1}=\hat{D_{1}}+\hat{D_{2}}$ and $D_{2}=i(\hat{D_{2}}-\hat{D_{1}})$ in Eq. \eqref{K48}, one can obtain
\begin{equation}
    F_{\infty-in}\propto -\beta(1-i|\hat{D_{2}}|^2 \sinh(i\beta)), \label{K50}
\end{equation}
\begin{equation}
  F_{\infty-out}\propto -\beta(1+i|\hat{D_{1}}|^2 \sinh(i\beta)).   \label{K51}
\end{equation}
Therefore, the reflection coefficient of the rotating BH with small $c_{4}$ reads 
\begin{equation}
    |R|=|\frac{F_{\infty-out}}{F_{\infty-in}}|=\left(\frac{1+i|\hat{D_{1}}|^2 \sinh(i\beta)}{1-i|\hat{D_{2}}|^2 \sinh(i\beta)}\right),  \label{K52}
\end{equation}
and the corresponding GF becomes
\begin{equation}
    \gamma=\frac{F_{hor}}{F_{\infty in}}=\frac{(\omega\Xi+(q^2+ak))|A|^2}{\beta\big(1+i|\hat{D_{1}}|^2 \sinh(i\beta)\big)}.  \label{K53}
\end{equation}
Since the result is almost the same as Eq.\eqref{ss39}, Moreover, no explicit well in the effective potential behavior in Fig.\ref{Fs2}, the interpretation for the stability in the presence of a magnetic field will be the same as the nonexistent one.  
\section{Semi-analytical Greybody Radiation} \label{sec5}
In this section, we shall follow the method, which was reviewed in \cite{Sakalli:2022xrb} (and references therein) to analyze the greybody radiation for both cases of $c_{s}=0$ and $c_{4}\neq 0$. 

The general semi-analytic bounds for GFs are given by \cite{isadd}%
\begin{equation}
\sigma_{l}\left(  \omega\right)  \geqslant\sec h^{2}\left\{  \int_{-\infty
}^{\infty}\wp dr_{\ast}\right\}  , \label{36}%
\end{equation}

where $\sigma_{l}$ represents the GF and $\wp$ is formulated as follows%
\begin{equation}
\wp=\frac{\sqrt{\left(  h^{\prime}\right)  ^{2}+\left[  \omega^{2}%
-V-h^{2}\right]  ^{2}}}{2h}, \label{37}%
\end{equation}

by which $h$ is a positive function that satisfies the following condition:
$h\left(  -\infty\right)  =h\left(  +\infty\right)  =\omega.$ Normally, one follows the method of replacing the $V$ parameter with the potential obtained in Eq. (\ref{33}) and then employs  the tortoise coordinate to evaluate the GF (\ref{36}). On the other hand, that method is not always the best course of action to take. In fact, this method also fails in our situation. So, to overcome this discrepancy,  we set $h=\sqrt{\omega^{2}-V}$ in Eq. (\ref{37}). This allows us to rewrite the expression for GF (\ref{36}) as%
\begin{equation}
\sigma_{l}\left(  \omega\right)  \geqslant\sec h^{2}\left\{  \frac{1}{2}%
\int_{-\infty}^{\infty}\left\vert \frac{h^{\prime}}{h}\right\vert dr_{\ast}\right\}  , \label{38}%
\end{equation}

which corresponds to%

\begin{equation}
\sigma_{l}\left(  \omega\right)  \geqslant\sec h^{2}\left\{  \ln\left(
\frac{h_{peak}}{h_{\infty}}\right)  \right\}  =\sec h^{2}\left\{  \ln\left(
\frac{\sqrt{\omega^{2}-V_{peak}}}{\omega}\right)  \right\}  . \label{39}%
\end{equation}

One can immediately observe that Eq. (\ref{39}) is valid for $\omega^{2}>V_{peak}$, the peak value of the potential \cite{Sakalli:2022xrb}. Besides, Eq. (\ref{39}) can be rewritten as
\begin{equation}
\sigma_{l}\left(  \omega\right)  \geqslant\frac{4\omega^{2}\left(  \omega
^{2}-V_{peak}\right)  }{\left(  2\omega^{2}-V_{peak}\right)  ^{2}}. \label{40}%
\end{equation}

\begin{figure}[h]
\centering\includegraphics[width=11cm,height=9cm]{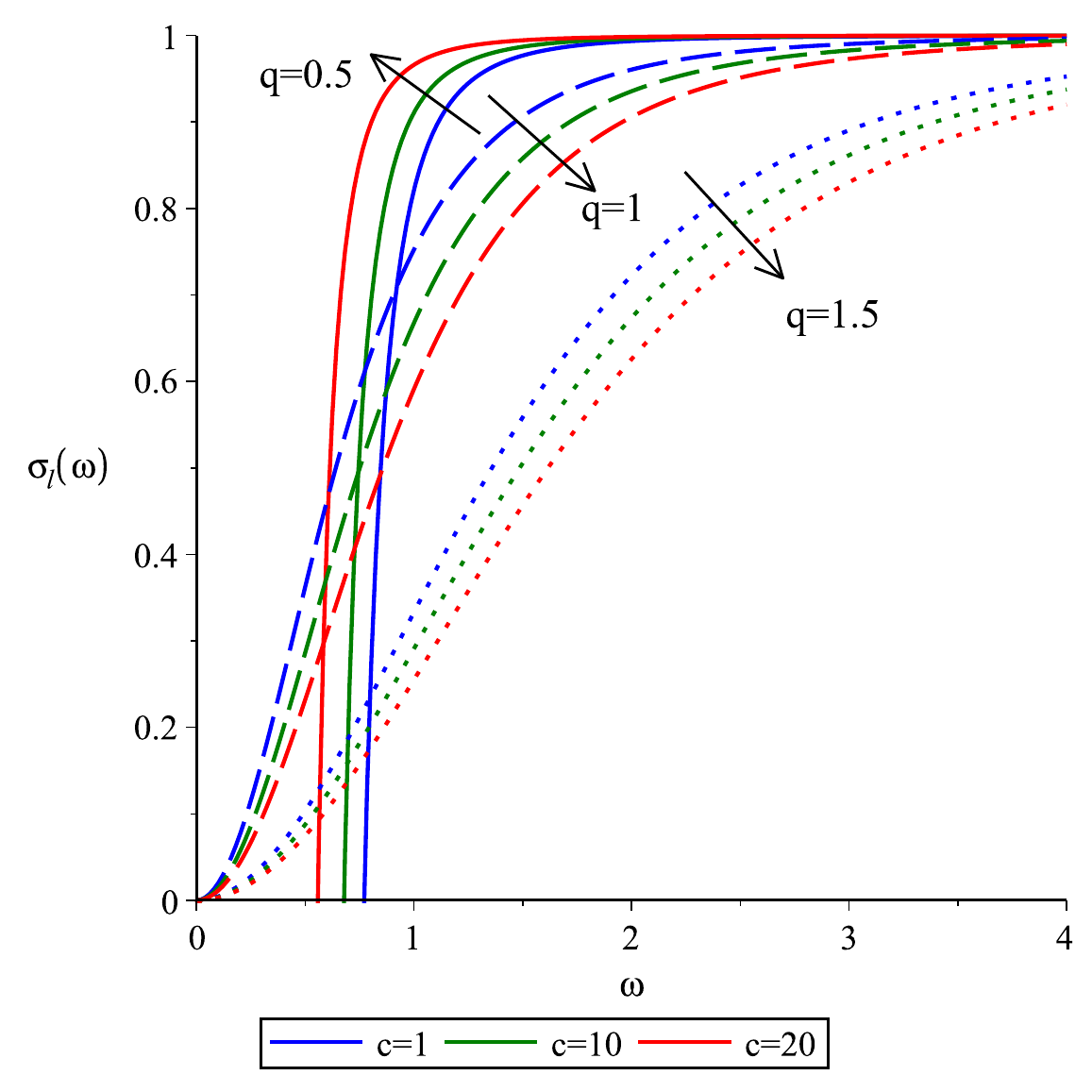}\caption{Plots of
$\sigma_{l}\left(  \omega\right)  $ versus $\omega$ for scalar particles.The
physical parameters are chosen as; $M=m=1,a=0.1$ and $\lambda=2$.} \label{Figure3}
\end{figure}

\begin{figure}[h]
\centering\includegraphics[width=11cm,height=9cm]{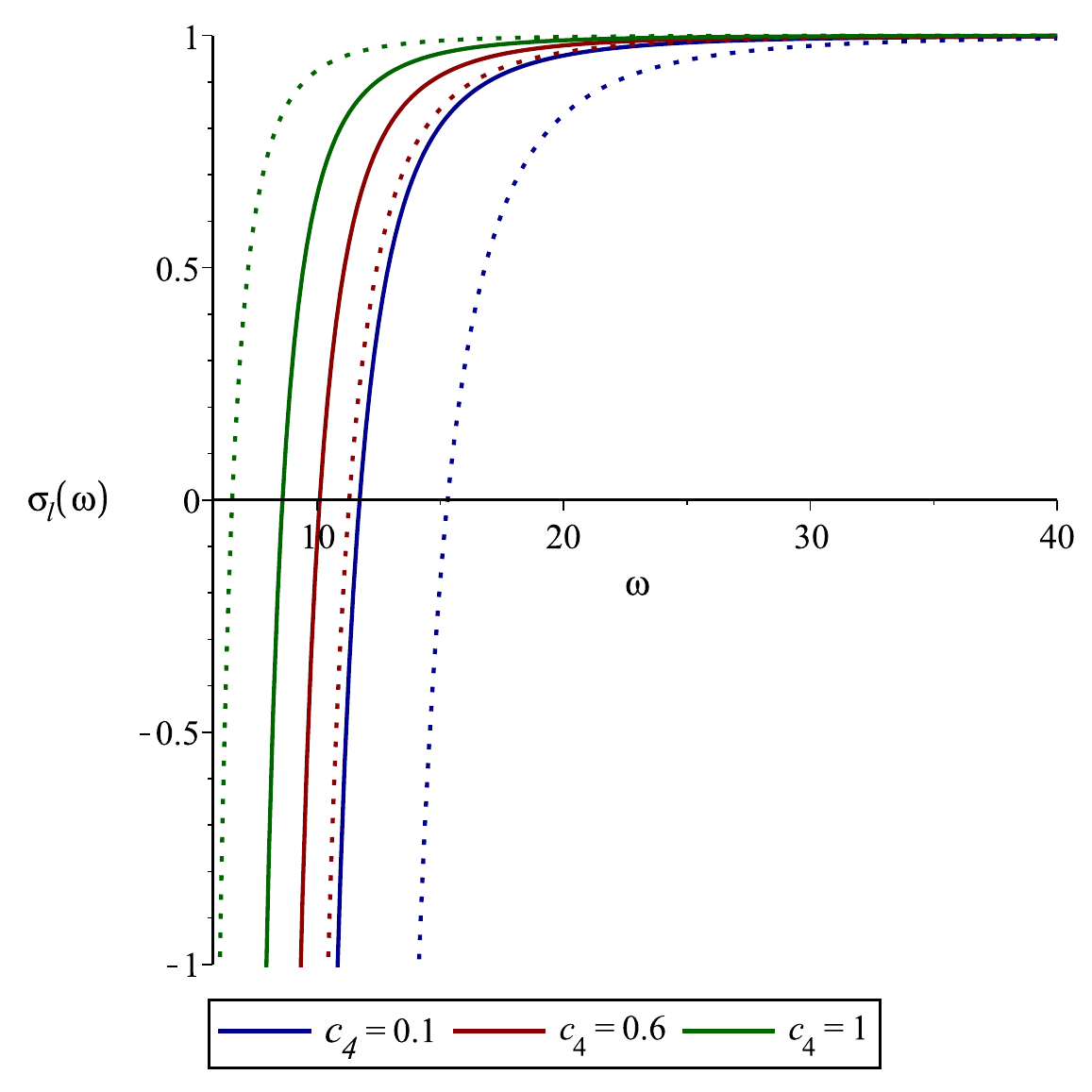}\caption{Plots of
$\sigma_{l}\left(  \omega\right)  $ versus $\omega$ for scalar particles. The dotted lines are represented for $q=3$ and solid lines for $q=2$. The physical parameters are chosen as; $M=m=k=1,a=0.1$, and $\lambda=2$.} \label{FigureS3}
\end{figure}

To find the maximum or the peak of the potential, \textcolor{blue}{first we derive the $r_{peak}$ from the effective potential equations by taking derivative with respect to $r$ }, as is well-known, one should find where the graph shifts from increasing to decreasing. To find out the rate at which the graph shifts from increasing to decreasing, we look at the second derivative and see when the value changes from positive to negative. Depending on the values of $V_{peak}$, the GFs are obtained. For instance, by setting $M=m=1$ and $\lambda=2$, in relation to the variables $q$ and $c$, the $V_{peak}$ expression is given by
\begin{equation}
V_{peak}\approx 0.00757c-\left(  0.08987c+2.0257\right)  q^{2}%
+1.18115-1.05389q^{4}. \label{41}%
\end{equation}
Substituting Eq. (\ref{41}) to Eq. (\ref{40}), we first compute the GFs of the rotating BH in non-linear Maxwell
$f\left(  R\right)$ gravity with $c_{4}=0$ for various parameters of $c$ and $q$. The behaviors of the obtained GFs for $c_{4}=0$ are illustrated in Fig. \ref{Figure3}. It is worth noting that although $c$ values should obey $0<c<2$ due to Ref. \cite{72}, we have used the values of $c$ above the relevant limit for the sake of revealing the changes in the GF behaviors. This by purpose exaggeration is made for just exhibiting the differences in the GF behaviors that are almost indistinguishable from each other within the theoretical limit of $0<c<2$. Fig. \ref{Figure3} shows a growth in the GF with $q=0.5$ by increasing the $c$ factor, but by increasing the charge value $q$ this course of action altered in the opposite direction. 
In a same circumstance, for infinitesimal $c_{4}$ case, $V_{peak}$ is found to be \begin{multline}
  V_{peak}\approx-A^2(100+62.80qc_{4})+0.2493046A(0.1243055-0.0309899q^2)+\\(10.150+q^2)^2+0.439065+0.054427q^2, \label{KS41} 
\end{multline}
where $A=0.501391+0.0621528q^2$. After substituting Eq. \eqref{KS41} in Eq. \eqref{40}, the obtained greybody radiation is depicted in Fig. \ref{FigureS3}, which shows that the greybody radiation explicitly increases with the magnetic field parameter $c_{4}$.    

\bigskip

\section{QNM\lowercase{s}} \label{sec6}

QNMs are important in the study of BH perturbation because they provide a way to understand the behavior of a BH in the presence of external perturbations such as a scalar, electromagnetic, gravitational, etc. When a BH is perturbed, it will respond by emitting gravitational waves that have a characteristic frequency. This frequency, known as the QNM, is determined by the properties of the BH, such as its mass, charge, and spin. QNMs of a BH is characterized by complex numbers. The complex frequency of a QNM is given by a real part and an imaginary part. The real part represents the oscillatory frequency of the mode, while the imaginary part represents the rate of decay or growth of the mode. By observing the QNMs in gravitational waves, astronomers can learn about the physical characteristics of the object that produced the waves.

In this section, for the scalar perturbations, we consider a semi-analytical approach to derive the frequencies of the QNMs of the charged rotating BHs
in nonlinear Maxwell $f\left(  R\right)  $ gravity. To this end, we employ the WKB (Wentzel-Kramers-Brillouin) approximation, which is a mathematical method used to solve differential equations (DEs) with a large-scale parameter. This approximation allows for a simplified solution to the differential equation and provides an estimate of the energy levels (frequency) within a certain accuracy. Conventionally, the WKB approach is based on the assumption that the solutions can be expressed as an exponential power series, where the coefficients of the series are determined by solving a set of recursive equations:
\begin{equation}
\Psi\left(  r\right)  \thickapprox\exp\left[  \frac{1}{\epsilon}%
\sum\limits_{n=0}^{\infty}\epsilon^{n}S_{n}\left(  r\right)  \right]  .\text{
\ \ \ \ }\epsilon\rightarrow0 \label{42}%
\end{equation}

The DE has a general form as%
\begin{equation}
\frac{d^{2}\Psi\left(  r\right)  }{dr^{2}}=Q\left(  r\right)  \Psi\left(
r\right)  , \label{43}%
\end{equation}

where $Q\left(  r\right)  =V(r)-\omega^{2}$, which contains two turning points. The boundary condition of waves is chosen to be $\Psi\left(  r\right)
=Z_{out}\Psi\left(  r\right)  _{out}$ for outgoing waves as $r_{\ast
}\rightarrow+\infty$ and $\Psi\left(  r\right)  =Z_{in}\Psi\left(  r\right)
_{in}$ for incoming waves while $r_{\ast}\rightarrow-\infty.$ First, Mashhoon \cite{73} invented this approach and applied it to the BHs in 1983. Then it was developed by \cite{74,75}.
The WKB approximation can be extended from the third to sixth order. The sixth-order WKB approximation, also known as the Konoplya approximation, is a method used to approximate the solutions of differential equations with complex potentials. The sixth-order WKB approximation by Konoplya can be found in his seminal papers \cite{67,Konoplya:2019hlu}. The Konoplya approximation uses a series expansion of the solution to the DE in powers of the small parameter $ \epsilon$, which is the wavelength of the solution.
Konoplya approximation for obtaining the complex frequencies of the QNMs is given by the following expression \cite{67}:
\begin{equation}
\omega^{2}=\left[  V_{0}+\sqrt{-2V_{0}^{\prime\prime}}\Lambda\left(  n\right)
-i\left(  n+\frac{1}{2}\right)  \sqrt{-2V_{0}^{\prime\prime}}\left(
1+\Omega\left(  n\right)  \right)  \right]  , \label{44}%
\end{equation}
where%
\begin{equation}
\Lambda\left(  n\right)  =\frac{1}{\sqrt{-2V_{0}^{\prime\prime}}}\left[
\frac{1}{8}\left(  \frac{V_{0}^{\left(  4\right)  }}{V_{0}^{\prime\prime
\prime}}\right)  \left(  \frac{1}{4}+\alpha^{2}\right)  -\frac{1}{288}\left(
\frac{V_{0}^{\prime\prime\prime}}{V_{0}^{\prime\prime}}\right)  ^{2}\left(
4+60\alpha^{2}\right)  \right], \label{45}%
\end{equation}
and%
\begin{multline}
\Omega\left(  n\right)  =\frac{1}{\left(  -2V^{\prime\prime}
_{0}\right)  }\left[  \frac{5}{6912}\left(  \frac{V_{0}^{\prime\prime\prime}%
}{V_{0}^{\prime\prime}}\right)  ^{4}\left(  77+188\alpha^{2}\right)  -\right.
\\
\left.  \frac{1}{384}\left(  \frac{V_{0}^{\prime\prime\prime}{}^{2}%
V_{0}^{\left(  4\right)  }}{V_{0}^{\prime\prime}{}^{\left(  3\right)  }%
}\right)  \left(  51+100\alpha^{2}\right)  +\frac{1}{2304}\left(  \frac
{V_{0}^{\left(  4\right)  }}{V_{0}^{\prime\prime}}\right)  ^{2}\left(
67+68\alpha^{2}\right)  +\right. \\
\left.  \frac{1}{288}\left(  \frac{V_{0}^{\prime\prime\prime}V_{0}^{\left(
5\right)  }}{V_{0}^{\prime\prime}{}^{\left(  2\right)  }}\right)  \left(
19+28\alpha^{2}\right)  -\frac{1}{288}\left(  \frac{V_{0}^{(6)}}{V_{0}%
^{\prime\prime}}\right)  \left(  5+4\alpha^{2}\right)  \right]. \label{46}%
\end{multline}

\begin{figure}[hbt!]
  \includegraphics[width=.49\textwidth]{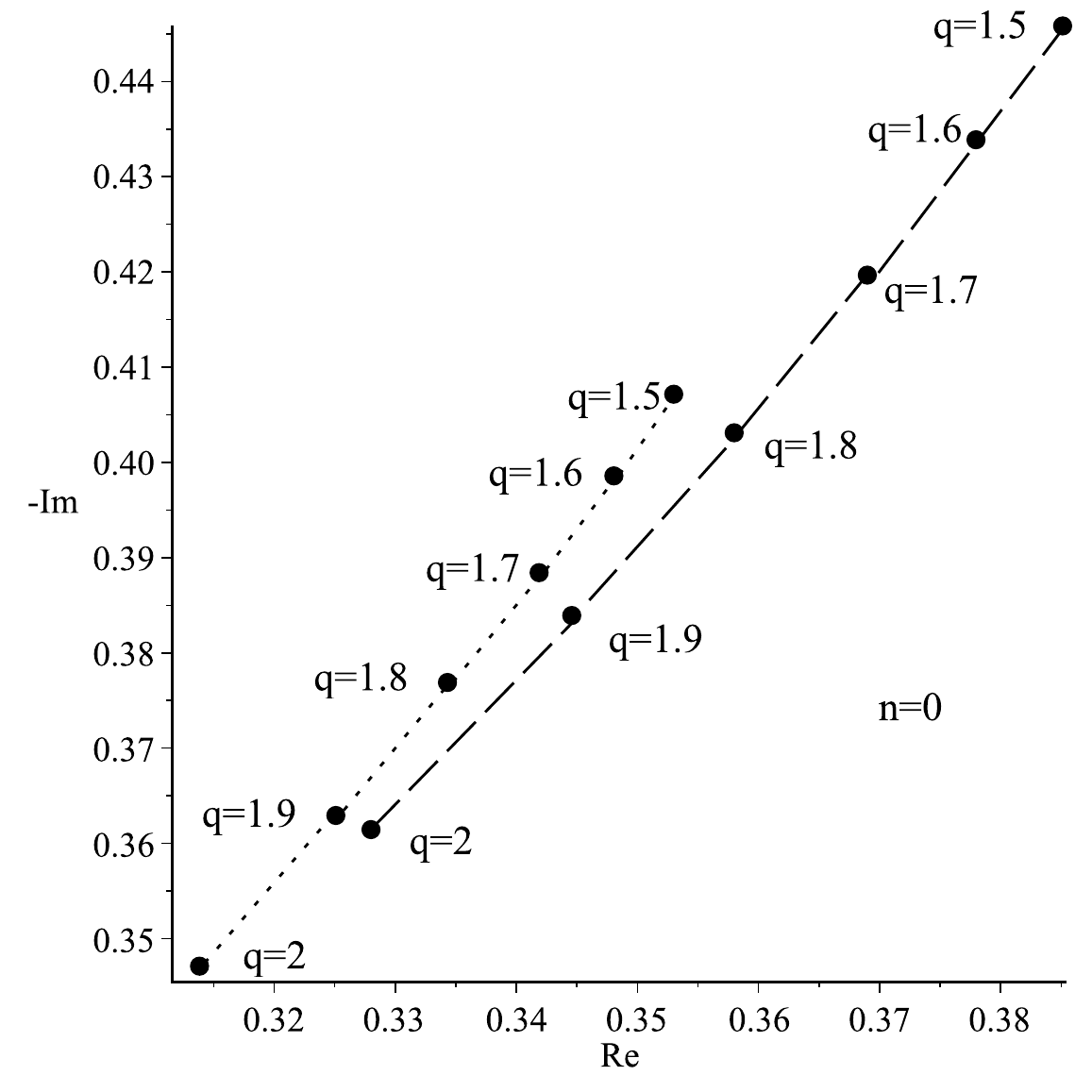}
  \includegraphics[width=.49\textwidth]{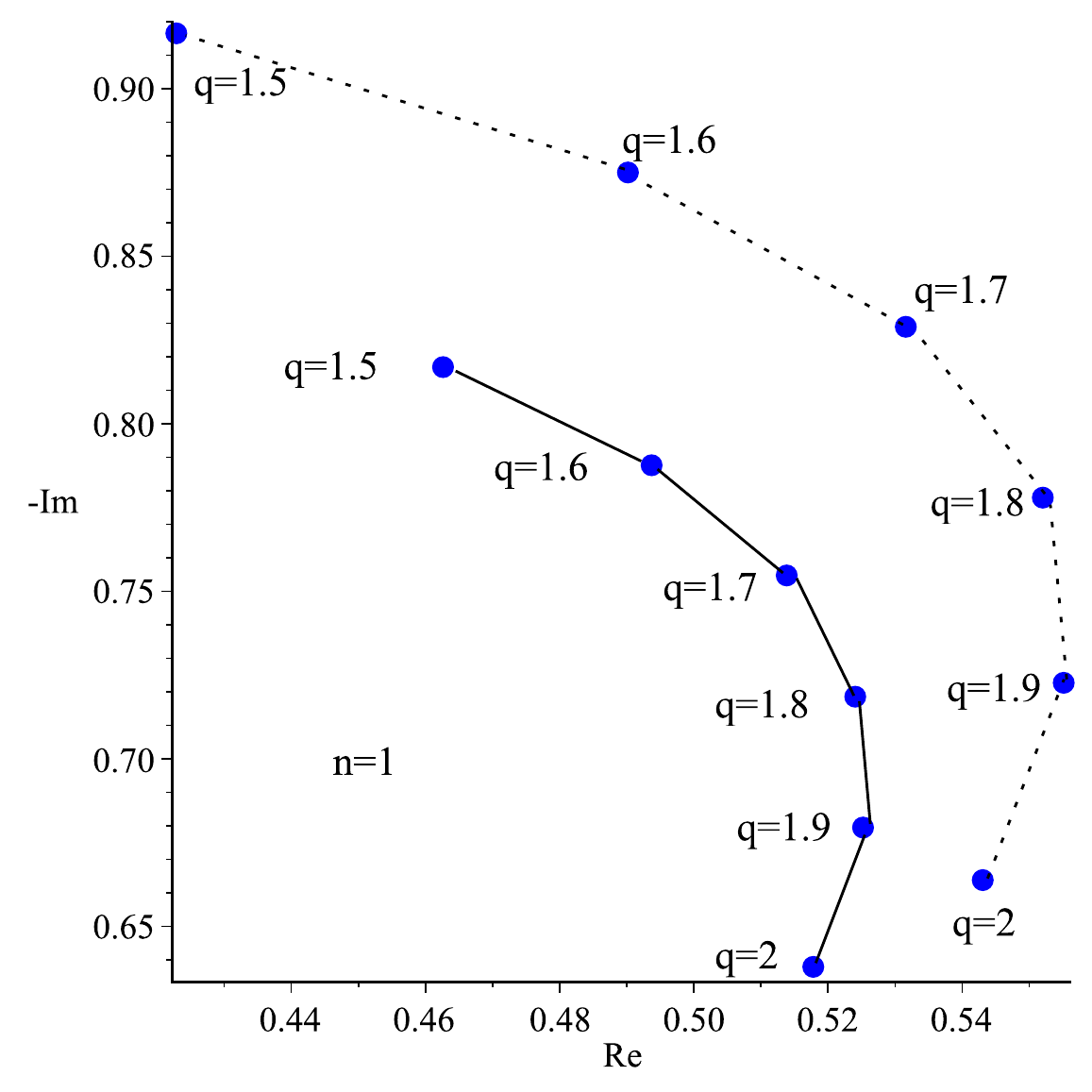}
\caption{Plots of QNMs of the rotating BH with $c_{4}=0$ under varying charge parameter $q$. The dotted line represents $c=1.9$ and dashed line stands for $c=1.7$; both for $n=0$ (left). The solid line is for $c=1.9$, however, the dotted line stands for $c=1.7$; both for $n=1$ (right).} \label{fig:14a}
\end{figure}

\begin{figure}[hbt!]
  \includegraphics[width=.49\textwidth]{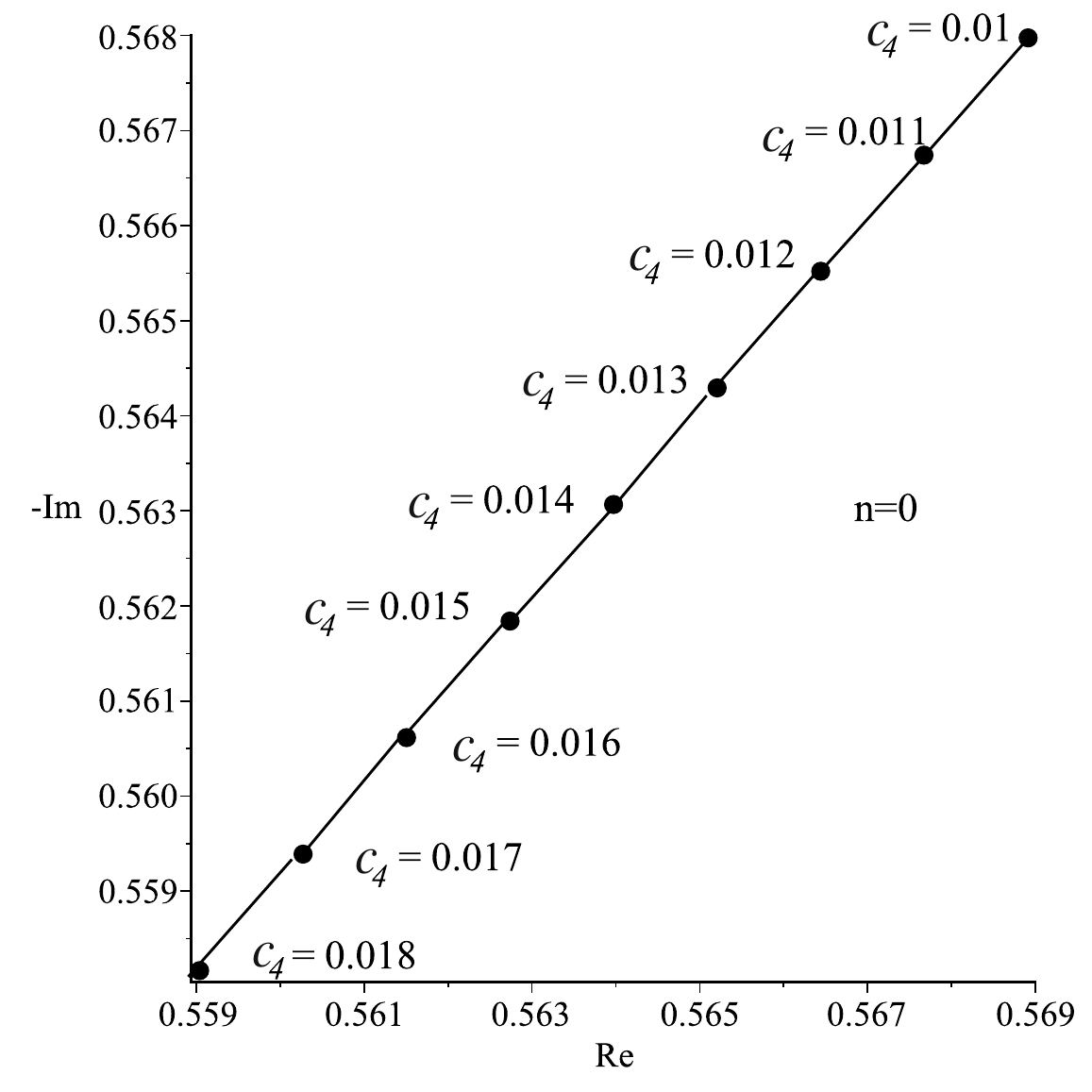}
  \includegraphics[width=.49\textwidth]{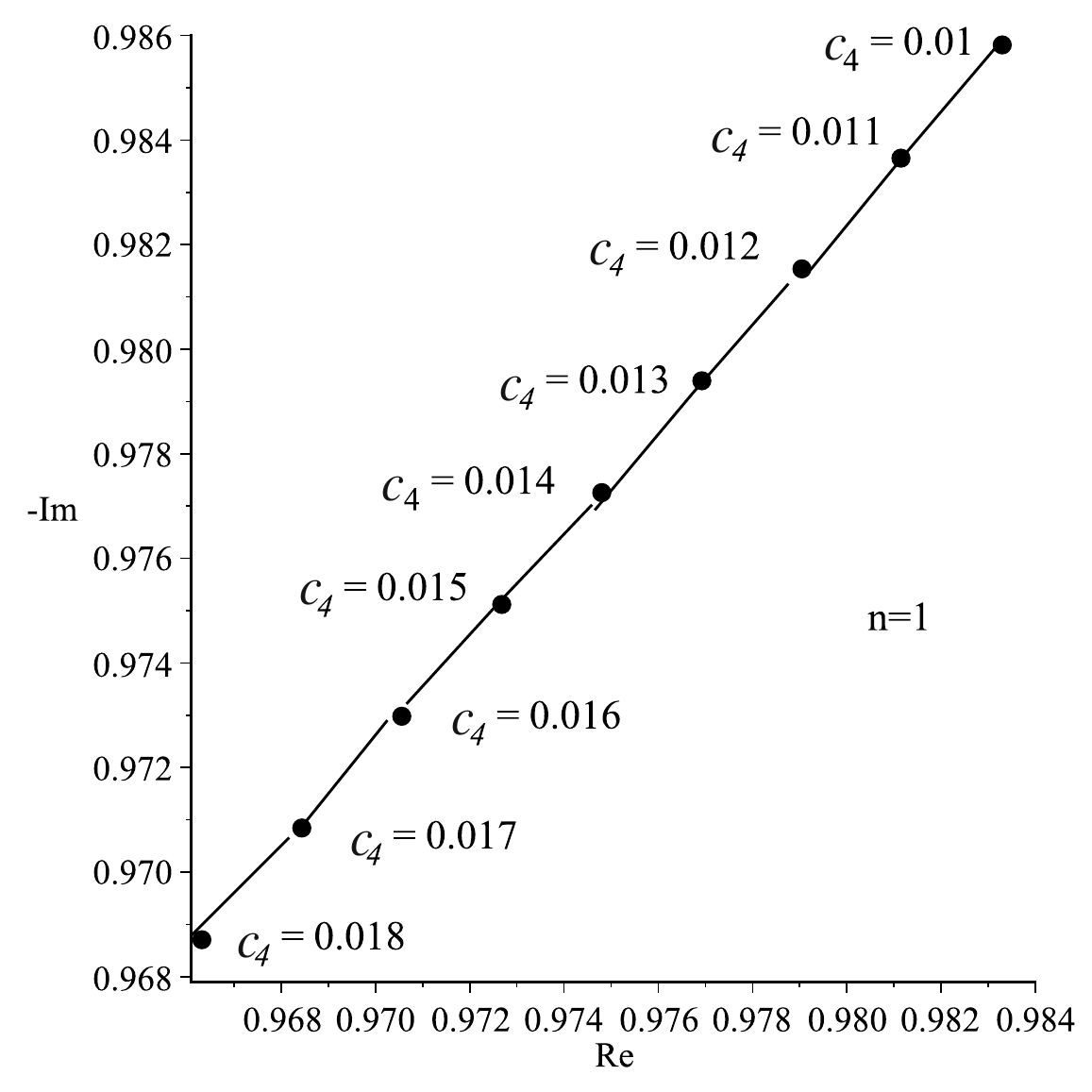}
\caption{Plots of QNMs of the rotating BH under varying $c_{4}$ values and fix charge $q=1$; for the tones of $n=0$ (left) and $n=1$ (right).} \label{fig:14B}
\end{figure}

\begin{figure}[hbt!]
  \includegraphics[width=.49\textwidth]{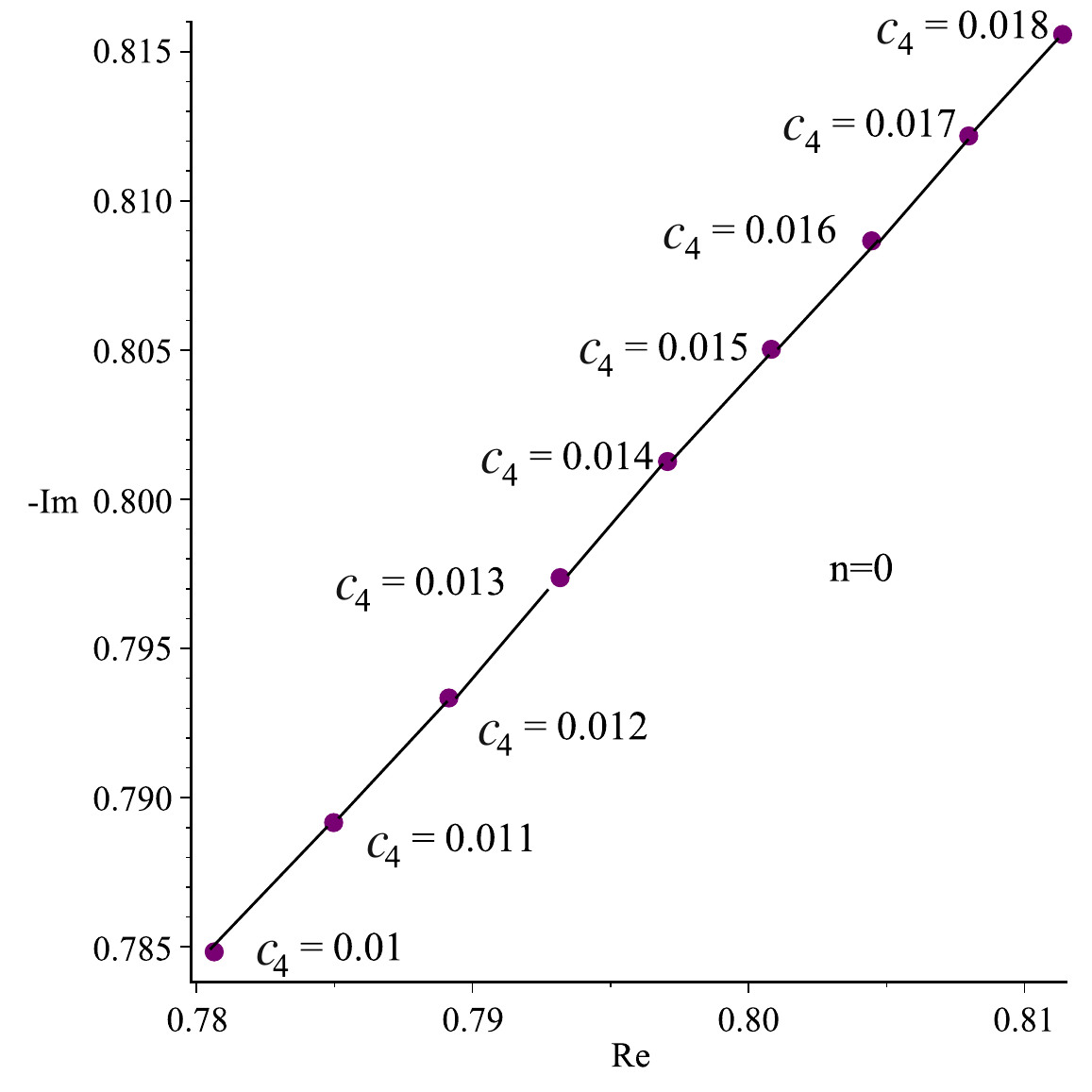}
  \includegraphics[width=.49\textwidth]{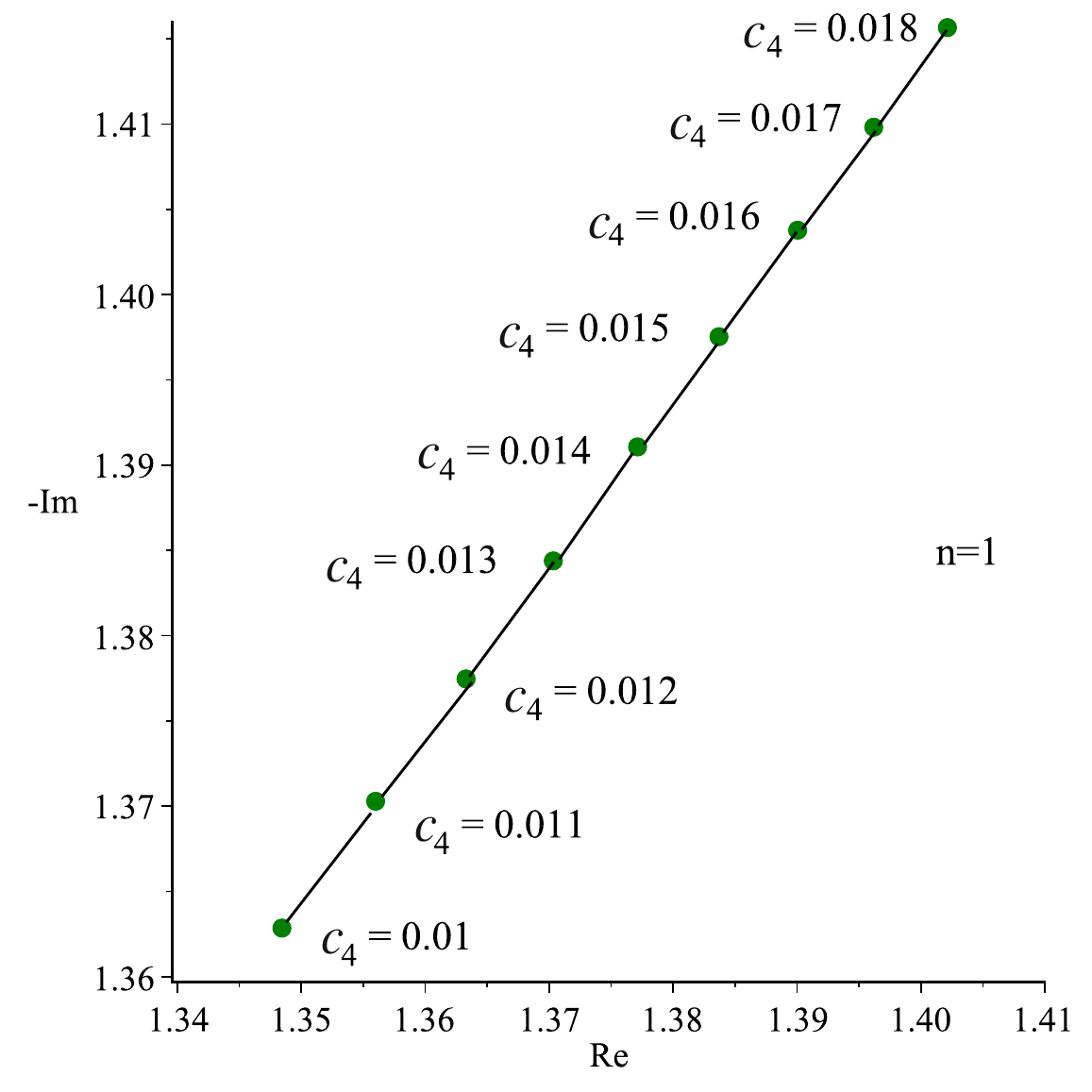}
\caption{Plots of QNMs of the rotating BH under varying $c_{4}$ values and fix charge $q=2$; for the tones of $n=0$ (left) and $n=1$ (right).} \label{fig:14C}
\end{figure}

In Eqs. (\ref{44})-(\ref{46}), the primes and superscripts ($4,5,6;$ for
the higher order derivatives) denote the differentiation with respect to
$r_{\ast}$ and $\alpha=n+\frac{1}{2}$, where $n$ denotes the tone number. By considering the effective potentials of both solutions, the results are tabulated in Tables \eqref{T1} and \eqref{T2} for the zero and non-zero magnetic field constants, respectively.\\
The behaviors of the QNMs for $c_{4}=0$ are illustrated in Figs. \ref{fig:14a}. One can observe that both parts (real and imaginary) of the QNMs decrease by increasing the charge parameter, $q$. Moreover, for $n=0$, QNMs increase by growing the $c$ parameter and then start to decrease. In addition, when $n=1$ and $q$ increases, the oscillation frequencies rise and the damping mode steadily declines.\\
On the other hand, for the case of $c_{4}\neq 0$, both parts of the QNMs decrease by increasing the magnetic field constant. The corresponding behaviors are depicted in Figs. \ref{fig:14B} and \ref{fig:14C}.

\begin{table}
   \centering
   \begin{tabular}{|c|c|c|c|c|c|c|c|}
\hline
$l$ & $n$ & $c$ & $q$ & $\omega_{Bosons}$ & $n$ & $q$ & $\omega_{Bosons}$\\
\hline
1 & 0 & 1.9 & 1.5 & 0.3531092772-0.4070157966i & 1 & 1.5 &
0.4628156990-0.8165375596i\\
&  &  & 1.6 & 0.3481646244-0.3984431959i &  & 1.6 &
0.4939092271-0.7872487864i\\
&  &  & 1.7 & 0.3419910035-0.3882796385i &  & 1.7 &
0.5140501121-0.7543542970i\\
&  &  & 1.8 & 0.3344221195-0.3764501224i &  & 1.8 &
0.5242699290-0.7181651739i\\
&  &  & 1.9 & 0.3252048163-0.3627849969i &  & 1.9 &
0.5254316095-0.6791172459i\\
&  &  & 2 & 0.3139459780-0.3469628546i &  & 2 & 0.5180125636-0.6375332598i\\
\hline
&  & 1.8 & 1.5 & 0.4171149837-0.4733886692i &  & 1.5 &
0.4883526041-0.9338811423i\\
&  &  & 1.6 & 0.4079913458-0.4593160345i &  & 1.6 &
0.5382395015-0.8962742366i\\
&  &  & 1.7 & 0.3965522968-0.4425772439i &  & 1.7 &
0.5699039647-0.8530833387i\\
&  &  & 1.8 & 0.3822146829-0.4227292265i &  & 1.8 &
0.5845200557-0.8033010248i\\
&  &  & 1.9 & 0.3639735647-0.3989634900i &  & 1.9 &
0.5823209373-0.7458662456i\\
&  &  & 2 & 0.3398557041-0.3697577443i &  & 2 & 0.5612257904-0.6787904390i\\
\hline
&  & 1.7 & 1.5 & 0.3852295784-0.445685725i &  & 1.5 &
0.4230520984-0.9161599009i\\
&  &  & 1.6 & 0.3780735086-0.4337333922i &  & 1.6 &
0.4903805959-0.8746300965i\\
&  &  & 1.7 & 0.3690969050-0.4195188351i &  & 1.7 &
0.5318064233-0.8285691459i\\
&  &  & 1.8 & 0.3581027910-0.4029669396i &  & 1.8 &
0.5522314188-0.7775780030i\\
&  &  & 1.9 & 0.3446888390-0.3837822213i &  & 1.9 &
0.5553476661-0.7223112093i\\
&  &  & 2 & 0.3281070990-0.3613035399i &  & 2 & 0.5432743870-0.6634326053i\\
\hline
\end{tabular}
  \caption{Bosonic QNMs of rotating BH in non-linear Maxwell $f(R)$ gravity for zero magnetic field parameter: $c_{4}=0$.} \label{T1}
\end{table}

\begin{table}
   \centering
   \begin{tabular}{|c|c|c|c|c|c|c|c|}
\hline
$l$ & $n$ & $q$ & $c_{4}$ & $\omega_{Bosons}$ & $n$ & $c_{4}$ & $\omega_{Bosons}$\\
\hline
1 & 0 & 1& 0.01 & 0.5689309522-0.5679607840i & 1 & 0.01 &
0.9833280391-0.9857933522i\\
&  &  & 0.011 & 0.5676889204-0.5667273000i &  & 0.011 &
0.9811788557-0.9836281427i\\
&  &  & 0.012 & 0.5664589707-0.5655057950i &  & 0.012 &
0.9790776944-0.9815108377i\\
&  &  & 0.013 & 0.5652237736-0.5642789170i &  & 0.013 &
0.9769537380-0.9793708001i\\
&  &  & 0.014 & 0.5639890661-0.5630524167i &  & 0.014 &
0.9748305274-0.9772315143i\\
&  &  & 0.015 & 0.5627548196-0.5618262697i &  & 0.015 & 0.9727080152-0.9750929361i\\
&  &  & 0.016 & 0.5615210087-0.5606004477i &  & 0.016 & 0.9705861561-0.9729550175i\\
&  &  & 0.017 & 0.5602875984-0.559374939322i &  & 0.017 & 0.9684648987-0.9708177181i\\
&  &  & 0.018 & 0.5590545741-0.5581496853i &  & 0.018 & 0.9663442042-0.9686809854i\\
\hline
&  & 2 & 0.01 & 0.7806931526-0.7847869409i & 1 & 0.01 &
1.348548830-1.362771574i\\
&  &  & 0.011 & 0.7850205043-0.7891179837i &  & 0.011 &
1.356111851-1.370209477i\\
&  &  & 0.012 & 0.7891946437-0.7932951529i &  & 0.012 &
1.363406512-1.377383600i\\
&  &  & 0.013 & 0.7932235382-0.7973263608i &  & 0.013 &
1.370446745-1.384307536i\\
&  &  & 0.014 & 0.7971145023-0.8012189424i &  & 0.014 &
1.377245378-1.390993848i\\
&  &  & 0.015 & 0.8008742877-0.8049796989i &  & 0.015 & 1.383814279-1.397454164i\\
&  &  & 0.016 & 0.8045091827-0.8086149183i &  & 0.016 &
1.390164462-1.403699260i\\
&  &  & 0.017 & 0.8080250159-0.8121304478i &  & 0.017 &
1.396306156-1.409739150i\\
&  &  & 0.018 & 0.8114272287-0.8155317201i &  & 0.018 &
1.402248881-1.415583157i\\
\hline
\end{tabular}
  \caption{Bosonic QNMs of rotating BH in non-linear Maxwell $f(R)$ gravity for infinitesimal magnetic field parameter $c_{4}.$}, \label{T2}
\end{table}

\section{Conclusion} \label{sec7}
In this paper, we have studied the superradiant instability/stability of a rotating BH in non-linear Maxwell $f(R)$ gravity under the influences of $q$, $c$, and $c_{4}$. To examine the (in)stability in this spacetime under the Dirichlet boundary condition, we have expanded the solutions into three different regions; near the event horizon (Region 1), intermediate (Region 2), and asymptotic (Region 3) regions. We next used the semi-analytic method outlined in Sec. \ref{sec5} to determine the GFs of the BH. To this end, we have followed a less problematic method and replaced the expression $h=\sqrt{\omega^{2}-V}$ with Eq. \eqref{37} to reach Eq. \eqref{39} or Eq. \eqref{40}. Then by defining the $V_{peak}$, the GFs of the rotating BH in non-linear Maxwell $f(R)$ gravity with/without magnetic field constant have been computed. The results obtained have been depicted in Fig. \ref{Figure3} to reveal the effects of $q$ and $c$ parameters on the GFs. The supreme point in the GFs behavior belongs to the $c=1$ and $q=0.5$, thereafter, by increasing the values of both $c$ and $q$ the GFs decrease.

To analyze the QNMs originated from the scalar perturbations of the rotating BH in non-linear Maxwell $f(R)$ gravity with/without magnetic field constant, we have considered the  $6{^{th}}$ order WKB approximation or the so-called Konoplya approximation. The results obtained have been both tabulated and illustrated in Figs. \ref{fig:14a} and \eqref{fig:14a}. Thus, we have shown the influence of $q$ and $c$ parameters on the QNMs. According to the relevant results, the QNMs with $n=0$ case have more stability than $n=1$ state. Similar to the real part of the QNMs, which decrease when the charge parameter is increased, the damping rate component (imaginary part) of the QNMs exhibits the almost same behavior, as illustrated in Figs. \ref{fig:14B} and \eqref{fig:14C}. In addition, the obtained QNMs have been presented in Tables \ref{T1} and \ref{T2} under different physical parameter changes and shown that the results support Figs. \ref{fig:14B} and \ref{fig:14C}.

In our findings, we have discovered that all unstable modes exhibit superradiance and all stable modes do not, consistent with the superradiant condition \eqref{ss40}. This means that scalar waves can experience superradiant amplification by extracting charge from the BH, indicating that the BH geometry is unstable. Additionally, superradiance can also be used to probe the fundamental principles of quantum gravity, such as the behavior of quantum particles in the presence of strong gravitational fields. Through the study of superradiance, researchers can gain a better understanding of the interplay between quantum mechanics and general relativity, leading to new insights into the nature of space, time, and gravity. It may be better understood if one considers black hole thermal fluctuations \cite{JHAP3}. And finally, we would like to state that we plan to extend this study to the fermionic perturbations of the rotating BH of the non-linear Maxwell $f(R)$ gravity in the near future.

\end{document}